\documentclass{article}

\usepackage[english]{babel}
\usepackage{amssymb}
\usepackage{authblk}

\usepackage[letterpaper,top=2cm,bottom=2cm,left=3cm,right=3cm,marginparwidth=1.75cm]{geometry}

\usepackage{amsmath}
\usepackage{graphicx}
\usepackage[colorlinks=true, allcolors=blue]{hyperref}

\title{LooPIN:  A PinFi protocol for decentralized computing}

\author[1]{Yunwei~Mao}
\author[1]{Qi~He}
\author[2,*]{Ju~Li}
\affil[1]{Loopro Inc., Cambridge, MA 02139, USA (\href{https://www.loopro.ai/}{https://www.loopro.ai/})}
\affil[2]{Massachusetts Institute of Technology, Cambridge, MA 02139, USA.}
\affil[*]{has never owned and will never own any LooPIN tokens.}
\date{November 15, 2025}

\begin{document}
\maketitle

\begin{abstract}
Networked computing power is a critical utility in the era of artificial intelligence. This paper presents a novel Physical Infrastructure Finance (PinFi) protocol designed to facilitate the distribution of computing power within networks in a decentralized manner. Addressing the core challenges of coordination, pricing, and liquidity in decentralized physical infrastructure networks (DePIN), the PinFi protocol introduces a distinctive dynamic pricing mechanism. It enables providers to allocate excess computing resources to a ``dissipative" PinFi liquidity pool, distinct from traditional DeFi liquidity pools, ensuring seamless access for clients at equitable, market-based prices. This approach significantly reduces the costs of accessing computing power, potentially to as low as 1\% compared to existing services, while simultaneously enhancing security and dependability. The PinFi protocol is poised to transform the dynamics of supply and demand in computing power networks, setting a new standard for efficiency and accessibility.
\end{abstract}

\section{Introduction}

The Artificial Intelligence (AI) landscape is experiencing a paradigm shift from centralized, proprietary models to decentralized, open-source systems, moving from dependence on trusted intermediaries to trust in algorithmically verified computations. Centralized services, despite their developer-friendly setup, often fall short of serving end-users' interests effectively. For instance, OpenAI's ChatGPT, despite its market dominance, has faced service disruptions due to its centralized architecture\footnote{See the uptime history at \url{https://status.openai.com/uptime}}. Our research highlights the benefits of decentralization, showing that migrating computing services, like the Llamma 70B model, to decentralized networks could drastically reduce downtime.

The shift towards decentralization, however, introduces challenges, notably higher deployment costs on decentralized platforms such as Akash Network, Nosana, io.net, and Render Network\footnote{Please refer to their websites,  \href{https://akash.network/}{Akash Network}, \href{https://nosana.io/}{Nosana}, \href{https://io.net/}{io.net}, \href{https://rendernetwork.com/}{Render Network}}. These costs stem not from a lack of computing resources but from centralized aspects of their pricing and liquidity models. Recognizing networked computing resources as an essential commodity underlines the need for a dedicated liquidity protocol to navigate the market's unique supply-demand dynamics.

Our project diverges from the conventional path of creating another decentralized computing network (DCN). We aim to introduce a liquidity protocol specifically designed for the decentralized computing market, addressing coordination, pricing, and liquidity challenges. Utilizing smart contracts and "dissipative" liquid pools our protocol intends to transform the decentralized computing field by improving efficiency and promoting fair resource distribution.

\begin{figure}[ht]
\centering
\includegraphics[width=0.65\linewidth]{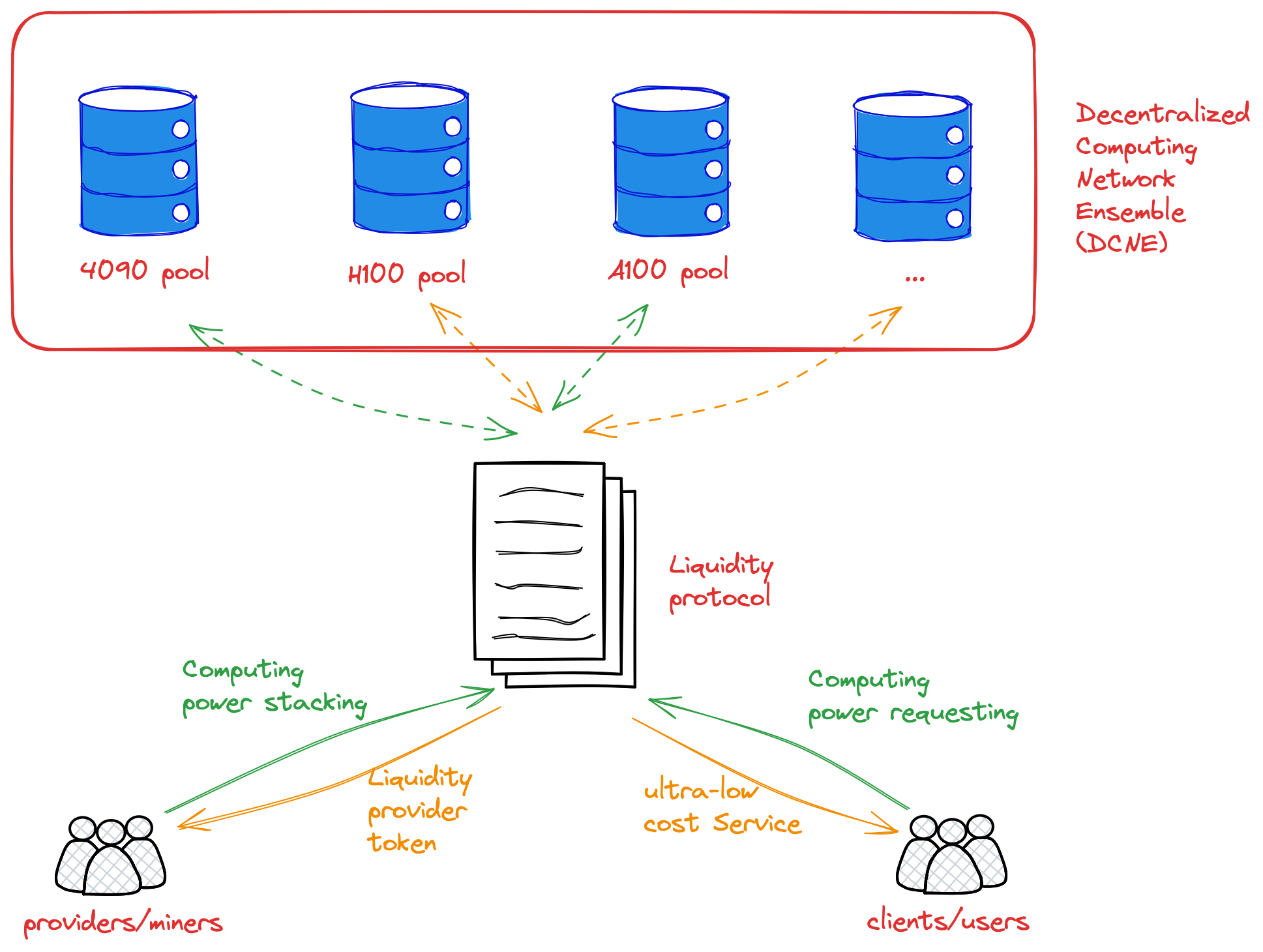}
\caption{\label{fig:scheme}The scheme of LooPIN PinFi system.}
\end{figure}

The LooPIN protocol, illustrated in Figure \ref{fig:scheme}, introduces a cutting-edge decentralized framework that enables seamless interactions between computing power providers and users. Built on three core rules, the protocol aims to facilitate a smooth exchange of computing resources in a secure and efficient environment:
\begin{itemize}
\item {\bf{Resource Staking}}: Providers, or miners, stake tokens as a commitment to contribute their computing resources to the network's liquidity pools, enhancing both the security and stability of these pools.

\item {\bf{Resource Maintenance and Utilization Rewards}}: For maintaining and making their computing resources available, providers receive tokens as compensation. Additional rewards are given when these resources are utilized by clients or developers, fostering a collaborative ecosystem.

\item {\bf{Resource Acquisition}}: Clients and developers can access the computing resources by contributing tokens to the liquidity pool, enabling them to execute tasks such as inference, fine-tuning, and training efficiently and transparently.
\end{itemize}
The design of the LooPIN protocol promotes a mutually beneficial environment, ensuring equitable and effective distribution of computing resources across the decentralized computing sector.

The remainder of this paper is organized as follows. Section \ref{overview} provides a comprehensive overview of the PinFi protocol's components, setting the foundation for further discussion. In Section \ref{pocps}, we delve into the Proof-of-Computing-Power-Staking (PoCPS) scheme, an innovative approach employed by the LooPIN network to cryptographically ensure the continuous provision of computing resources as claimed by miners. Section \ref{liquidpool} explores the liquidity pool in detail, with a particular focus on its unique dissipative nature. Transaction types within our system are systematically defined in Section \ref{txs}, establishing a clear understanding of the operational framework. The paper concludes in Section \ref{ongoingwork} with reflections on potential future developments and avenues for research, positioning our work within the broader context of the evolving decentralized computing landscape.

~\\~\\
{\bf{Note}}: This manuscript is a living document, reflecting our commitment to transparency and adaptation in the face of evolving technological landscapes. As we progress through the development of the LooPIN protocol, updates and modifications to this document will be made to mirror the latest advancements and iterations within our project. 
%It is important to note that due to the iterative nature of our development process, there may be discrepancies between the current implementation and the descriptions provided herein.

We warmly encourage readers with an interest in our work to explore our GitHub repository at \url{https://github.com/loopin-project}. Our commitment to open-source principles means we will progressively share various components of the LooPIN system, inviting collaboration, feedback, and community participation.

\section{Components of the LooPIN PinFi Protocol}
\label{overview}
This section outlines the LooPIN network, detailing its key participants, foundational components, and overarching protocol.

\subsection{Key Participants}
In the LooPIN protocol ecosystem, interaction and functionality revolve around four distinct types of participants: Devices, Miners, Verifiers, and Clients. Each plays a critical role in maintaining and utilizing the decentralized computing network.
\begin{itemize}

\item {\bf{Devices}}. Devices represent the foundational physical computing resources within the liquid pools of the decentralized computing network ensemble (DCNE). Critical information pertaining to these Devices—including SSH details, environment setup, and availability—is uniquely fingerprinted. This digital fingerprint is then securely stored on the blockchain, ensuring transparency and verifiability.

\item {\bf{Miners}}. Miners are users who contribute to the LooPIN network by providing Devices and staking the token to a designated liquidity pool. Through participating in the Proof-of-Computing-Power-Staking process, Miners validate their continuous contribution of computing power, available for Client use. At predetermined epochs, Miners are awarded block rewards, proportional to their contribution and the protocol's governance.

\item {\bf{Verifiers}}. Integral to maintaining the integrity of the LooPIN network, verifiers play a crucial role in assessing the operational status of devices. Leveraging the Proof-of-Computing-Power-Staking scheme, verifiers are empowered to efficiently validate that devices not only claim to, but actually do, provide secure and reliable services to clients. 

\item {\bf{Clients}}. Clients engage with the LooPIN network by utilizing computing resources from specific liquidity pools. Similar to Miners, Clients are eligible for block rewards at designated epochs, acknowledging their participation and contribution to the network's computational demands.
\end{itemize}

\subsection{Elementary Components of the LooPIN Network}
The LooPIN network is underpinned by several critical components, each contributing to its innovative and decentralized architecture:

\begin{itemize}
\item {\bf{Liquidity Pools}}: Central to our network are the liquidity pools, differentiated by the type of computing power resource they represent (e.g., GPUs, and potentially in the future, TPUs or Groq chips). These pools are designed to be permissionless and resistant to censorship, enabling anyone to establish a liquidity pool on the blockchain via our smart contracts. The liquidity pools within the LooPIN protocol exhibit a dissipative nature, a characteristic that we will explore in greater depth in Section \ref{liquidpool}.
\item {\bf{Decentralized Computing Network Ensemble (DCNE)}}: The aggregation of these permissionless liquidity pools, contributed by a variety of independent computing providers, forms what we refer to as the Decentralized Computing Network Ensemble (DCNE). This can be thought of as a 'network of networks', representing a new layer of decentralized computing infrastructure.
\item {\bf{Proof-of-Computing-Power-Staking (PoCPS)}}: We introduce an innovative, computationally efficient method for miners to validate their contribution of computing resources to the network. Through a smart contract, miners can demonstrate their provision of computing power in a manner that is both cryptographically secure and relative in time compared to others on the network.
\end{itemize}

\subsection{Protocol Overview}
Integrating Proof-of-Computing-Power-Staking (PoCPS) within a blockchain framework, the LooPIN protocol leverages:

\begin{itemize}
\item {\bf{Foundational Architecture}}: At its core, LooPIN is a decentralized network of computing resources leveraging the innovative PoCPS algorithm. This is supported by a blockchain infrastructure that utilizes a native token, facilitating secure and transparent transactions within the network.
\item {\bf{Dynamic Pricing through Automatic Market Making}}: The protocol employs an automatic market-maker model to manage the staking of computing power and the fulfillment of computing requests. This model allows for dynamic pricing, with the cost of computing power fluctuating based on market supply and demand.
\item {\bf{Token Rewards}}: Miners are rewarded with tokens for committing their computing resources to the network's liquidity pools. Similarly, clients are compensated for using this staked computing power for AI-related tasks. This system incentivizes participation and ensures a balanced distribution of rewards and resources within the network.
\end{itemize}

\section{Proof-of-Computing-Power-Staking}
\label{pocps}
Within the LooPIN network, a foundational requirement for miners is to authenticate their contribution of computing power to the Decentralized Computing Network Ensemble (DCNE). This authentication is facilitated by adherence to the Proof-of-Computation-Power-Staking (PoCPS) protocol. The network, along with designated verifiers, conducts audits to validate these claims, ensuring the integrity of the staking process. The primary aim of the PoCPS protocol is to verify that miners accurately represent their computational contributions in comparison to others within the network. It also establishes a framework to cryptographically identify and prove any instances of dishonesty.

This section delves into the motivations behind the PoCPS scheme, its conceptual framework, and the practical aspects of its implementation within the LooPIN ecosystem. Our goal is to articulate the importance of PoCPS in maintaining a transparent, secure, and trustworthy environment for all network participants, ensuring that miners are held to the highest standards of integrity in their contributions to the DCNE.

\subsection{Motivation}

The LooPIN network employs stringent security protocols to safeguard against three principal attack vectors that could potentially be exploited by malicious actors to gain undue advantages: Sybil attacks, outsourcing attacks, and lack-of-commitment attacks. 

\begin{enumerate}

\item {\bf{Sybil Attacks}}: These attacks involve miners utilizing virtualization techniques to falsely inflate their computing capabilities, thereby claiming unjust compensation. This undermines the fair allocation of resources and rewards within the network.
\item {\bf{Outsourcing Attacks}}: Miners might promise more computing power than they physically possess, intending to compensate for this shortfall by temporarily utilizing resources from the network’s liquidity pools. This misleads the network about its actual resource capacity and jeopardizes the reliability of services.
\item {\bf{Lack-of-commitment Attacks}}: In these scenarios, miners may falsely assert that they are dedicating their computing resources exclusively to a specific liquidity pool, when in reality, they fail to do so. This lack of commitment can lead to inconsistencies in resource availability and allocation.
\end{enumerate}

Proof-of-Computing-Power-Staking (PoCPS) emerges as a pioneering verification mechanism, drawing inspiration from the Proof-of-Replication (PoRep) utilized within the Filecoin system\cite{filecoin2017}. Studies have demonstrated that PoRep-based proofs can effectively mitigate three distinct types of attacks, highlighting their robustness in ensuring data integrity and security\cite{filecoin2017wip}. Nevertheless, PoRep, rooted in Proof-of-Storage concepts\cite{ateniese2007provable}, is optimally functional in scenarios involving data processing and/or storage. This attribute limits its applicability in environments demanding intensive GPU resources, as is the case with our computing network.

Addressing this limitation, PoCPS is tailored explicitly for GPU-intensive tasks within computing networks. By focusing on the unique demands of GPU-based operations, PoCPS offers a specialized solution that bridges the gap left by traditional Proof-of-Storage schemes. This innovation not only enhances the security framework for our GPU-centric computing environment but also ensures a greater degree of reliability and efficiency in verifying the contribution and utilization of computing resources.

\subsection{Constructing PoCPS}

\begin{figure}[ht]
\centering
\includegraphics[width=1.0\linewidth]{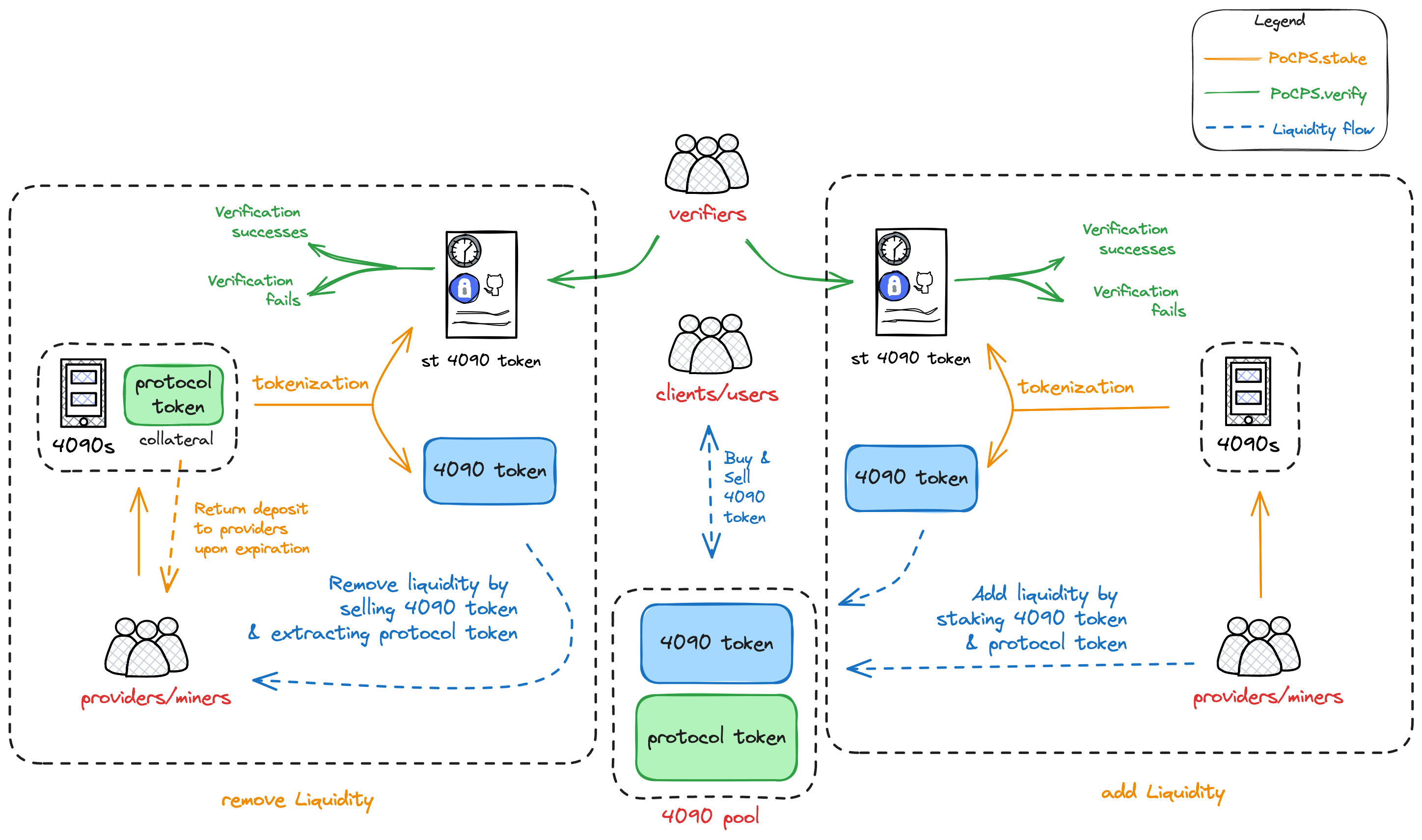}
\caption{\label{fig:example}An illustration of the scheme of PoCPS.}
\end{figure}

To elucidate the workings of Proof-of-Computing-Power-Staking (PoCPS), consider the example detailed in Figure \ref{fig:example}, involving a miner with an NVIDIA RTX 4090 GPU who stakes this asset in the LooPIN Decentralized Computing Network Ensemble (DCNE) for two hours. The staking mechanism unfolds in the following steps:

\begin{enumerate}

\item {\bf{Token Burn and Certificate Issuance}}: Miners destroy a designated quantity of tokens to receive a burn token certificate, marking their initial commitment to the network.

\item {\bf{Certificate Deposit and Token Allocation}}: By depositing this certificate along with a maintenance fee (waived for miners opting to be liquidity providers) into a smart contract address, miners are allocated two tokens: the st-4090-token, a non-tradable, non-fungible token acting as proof of liquidity provision with a 2-hour validity, and the 4090-token, whose existence is tied to the st-4090-token's lifespan.

\item {\bf{Liquidity Injection}}: Contributing the 4090-token and an equivalent utility token amount enhances the 4090 pool's liquidity. At the st-4090-token's expiration, two key events occur: the miner receives rewards for staking computing resources, as detailed in Section \ref{liquidpool}, and utility tokens are automatically returned to the miner at the prevailing exchange rate.

\item {\bf{Liquidity Removal}}: Miners may opt to sell their 4090-token to the pool for utility tokens, incurring an exchange fee. Upon expiration of the st-4090-token—and pending successful interactive verification—the maintenance fee is refunded.
\end{enumerate}

The PoCPS verification sequence is as follows:
\begin{enumerate}
\item {\bf{Selection and Challenge}}: Verifiers randomly select a st-4090-token and issue a challenge to its server, earning block rewards for their verification efforts.

\item {\bf{Outcome Determination}}: If a server (for liquidity removal miners) fails the challenge, the st-4090-token is immediately invalidated, and the maintenance deposit is burned. For liquidity provider miners facing a failed challenge, the st-4090-token expires, and the corresponding utility tokens are withdrawn from the pool and returned. A successful challenge maintains the st-4090-token's validity until the next verification point.
\end{enumerate}

Building on the illustrative example provided, PoCPS can be formed as follows. This framework draws inspiration from Proof-of-Replication (PoRep), sharing terminological and conceptual similarities. In maintaining brevity and clarity within this definition, we aim to outline the essential mathematical underpinnings of PoCPS succinctly. For readers interested in a more comprehensive exploration of the subject, we recommend referring to the detailed study available in the literature on PoRep, specifically \cite{filecoin2017wip}.

~\\
{\bf{Definition}} (Proof-of-Computing-Power-Staking) A PoCPS protocol enables an efficient server $\mathcal{S}$ to convince a challenger $\mathcal{C}$ that $\mathcal{S}$ is providing a physically unique and dedicated computing resource to the liquid pool. A PoCPS protocol is characterized by a tuple of polynominal-time algorithms: (Setup, Prove, Verify)
\begin{itemize}
\item $\mathcal{P}^\mathcal{S}, \mathcal{V}_\mathcal{S}, \mathcal{V}_\mathcal{C}\leftarrow \text{PoCPS.Setup}(\lambda, \mathcal{S})$, where $\mathcal{V}_\mathcal{S}$ and $\mathcal{V}_\mathcal{C}$ are scheme-specific setup variables for $\mathcal{S}$ and $\mathcal{C}$ respectively, that depend on a security parameter $\lambda$ and the computing power $\mathcal{S}$ that the miner has claimed. $\text{PoCPS.Setup}$ is used to initialize the proving scheme and give $\mathcal{S}$ and $\mathcal{C}$ information they will use to run $\text{PoCPS.Prove}$ and $\text{PoCPS.Verify}$.

\item $\pi^c \leftarrow \text{PoCPS.Prove}(\mathcal{V}_\mathcal{S}, \mathcal{P}^\mathcal{S}, c)$, where $c$ is a challenge, and $\pi^c$ is a proof that a server $\mathcal{S}$ has access to $\mathcal{P}^\mathcal{S}$ a specific computing power claimed by server $\mathcal{S}$. $\text{PoCPS.Prove}$ is run by $\mathcal{S}$ to produce a $\pi^c$ for challenger $\mathcal{C}$.

\item $\{0, 1\}\leftarrow \text{PoCPS.Verify}(\mathcal{V}_\mathcal{C}, c, \pi^c)$ which checks whether a proof is correct. $\text{Verify}$ is run by $\mathcal{C}$ and convinces $\mathcal{C}$ whether $\mathcal{S}$ indeed provides the amount of computing power $\mathcal{P}^\mathcal{S}$. 
\end{itemize}

In developing a secure Proof-of-Computing-Power-Staking (PoCPS) protocol, it is crucial to devise strategies that effectively mitigate Sybil Attacks, Outsourcing Attacks, and Lack-of-commitment Attacks. Our proposed construction for a time-bounded PoCPS is grounded in leveraging local time from the challenger's perspective, defining the proving scheme as time-bounded to ensure the validity of a proof is confined within a specific timeframe. This approach inherently means that if the prover $\mathcal{S}$ fails to generate a valid proof $\pi$ within the designated period following the receipt of challenge $c$, the proof becomes invalid due to the ample time potentially used to fabricate it.

For instance, in combatting the Sybil Attack (and similarly Lack-of-commitment Attacks), where the integrity and independence of physical computing devices must be verified to prevent the inflation of their numbers through virtualization, our strategy involves distinctly treating each physical device. This method requires a server $\mathcal{S}$ to commit in advance to a specific encoding result derived from a GPU-intensive Proof-of-Work (PoW) scheme, utilizing a unique per-device key $ek$. Formally, we express the result of the GPU-intensive PoW scheme as $\mathcal{P}^\mathcal{S}_{ek} = \text{PoW}(ek)$, ensuring that the output is unique for each device ($\mathcal{P}^\mathcal{S}_{ek_j}\neq \mathcal{P}^\mathcal{S}_{ek_i}$ for ${ek_j}\neq {ek_i}$). This uniqueness is crucial for the verification process, and to facilitate recovery of the key $ek$, the PoW scheme must be reversible, denoted as $ ek = \text{PoW}^{-1}(\mathcal{P}^\mathcal{S}_{ek})$.

In addressing Outsourcing Attacks within the PoCPS framework, the protocol must prevent a server from generating $\mathcal{P}^\mathcal{S}_{ek}$ on the fly — that is, between receiving the challenge $c$ and producing the proof $\pi^c$. The goal is to design the system such that an attacker attempting to produce $\mathcal{P}^\mathcal{S}_{ek}$ just-in-time would be significantly slower than an honest server responding to a challenge. Given that generating $\mathcal{P}^\mathcal{S}_{ek} = \text{POW}(ek)$ requires a distinguishable amount of time, the challenger $\mathcal{C}$ can differentiate between the response times of an honest server and an attacker based on the total time taken to return the proof, which includes both the round-trip time ($\text{RTT}^{\mathcal{C}\rightarrow \mathcal{S}\rightarrow \mathcal{C}}$) and the time required to prove possession of $\mathcal{P}^\mathcal{S}_{ek}$ or to produce it via PoW. Specifically, the total time $\mathcal{T}^{\text{\tiny honest}}$ for an honest response includes the RTT and the time to execute PoCPS.Prove using an already prepared $\mathcal{P}^\mathcal{S}_{ek}$, i.e.,
\begin{equation*}
\mathcal{T}^{\text{\tiny honest}} = \text{RTT}^{\mathcal{C}\rightarrow \mathcal{S}\rightarrow \mathcal{C}} + \text{Time}(\text{PoCPS.Prove}(\mathcal{V}_\mathcal{S}, \mathcal{P}^\mathcal{S}_{ek}, c))
\end{equation*}
while $\mathcal{T}^{\text{\tiny attack}}$ for an attack includes the RTT and the time to execute PoCPS.Prove while generating $\mathcal{P}^\mathcal{S}_{ek}$ through PoW on demand, i.e., 
\begin{equation*}
\mathcal{T}^{\text{\tiny attack}} = \text{RTT}^{\mathcal{C}\rightarrow \mathcal{S}\rightarrow \mathcal{C}} + \text{Time}(\text{PoCPS.Prove}(\mathcal{V}_\mathcal{S}, \text{PoW}(ek), c)).
\end{equation*}
The inherent delay in producing $\mathcal{P}^\mathcal{S}_{ek}$ for attackers ensures that $\mathcal{T}^{\text{\tiny attack}} \gg \mathcal{T}^{\text{\tiny honest}}$, making it possible for the challenger to identify and mitigate outsourcing attacks effectively.

\section{Liquidity Pools and Their Dissipative Dynamics}
\label{liquidpool}

This section delves into the liquidity pools within the LooPIN protocol, focusing on their intrinsic dissipative characteristics and the strategies implemented to navigate the challenges presented by the temporal depreciation of computing power hours.

\subsection{Source of the dissipation}
In traditional decentralized finance (DeFi) environments\cite{adams2020uniswap}, liquidity pools comprising token pairs, such as A and B, exhibit a static quality in the absence of external interactions. This means that the token quantities within these pools are not influenced by the passage of time, $t$, remaining constant unless affected by transactions or other forms of engagement. The invariant nature of these token pairs is a hallmark of conventional DeFi liquidity pools, where the absence of activity does not alter the balance of assets. 

\begin{figure}[ht]
\centering
\includegraphics[width=1.0\linewidth]{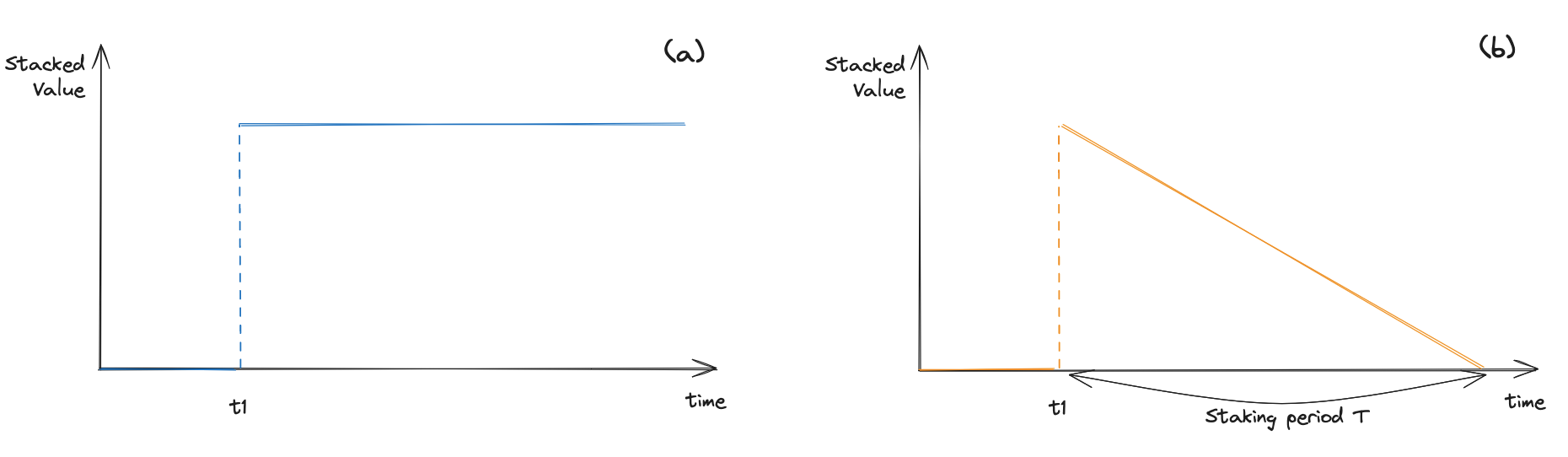}
\caption{\label{fig:dissipation}The conservative principles underlying decentralized finance (DeFi) systems (a), juxtaposed with the inherently dissipative dynamics of physical infrastructure finance (PinFi) systems (b).}
\end{figure}

In analyzing the liquidity injection within a decentralized finance (DeFi) system, illustrated in Figure \ref{fig:dissipation}(a), we observe a scenario where, at a specific time $t_1$, a value $\mathbb{S}$ of token A is contributed to a liquid pool. The value staked over time, denoted as $\mathcal{S}_{\text{\tiny DeFi}}$, can be modeled as:
\begin{equation}
\mathcal{S}_{\text{\tiny DeFi}} = \mathbb{S}\cdot u(t-t_i),
\end{equation}
where $u(\cdot)$ is the step function. Through this model, we derive the dissipation rate:
\begin{equation}
\mathcal{D}_{\text{\tiny DeFi}} = \mathbb{S}\cdot \delta(t-t_i).
\end{equation}
Here $\delta(\cdot)$ is the delta function, indicating that the dissipation rate is nonzero solely at the moment $t\equiv t_1$, the point of liquidity injection. Consequently, outside of this singular instance, the dissipation rate is zero, affirming that DeFi systems exhibit a conservative nature with no ongoing dissipation of the injected liquidity.

Contrastingly, the PinFi protocol introduces a paradigm shift with its innovative approach to liquidity pools, distinguishing itself from the traditional models prevalent in conventional DeFi. Within PinFi's framework, one of the tokens in the liquidity pool represents the total staked computing power hours—a dynamic and dissipative asset whose value inherently declines over time, $t$, reflecting the consumptive nature of computing resources. The other token in the pool is the protocol token, serving as the stable counterpart in this pair. Considering a simplified liquidity injection event for a PinFi system, shown in the Fig. \ref{fig:dissipation}(b). At time $t_1$, a value $\mathbb{S}$ of token A (computing power hour) is added to a liquid pool. The miner commits the computing power to the liquidity pool for $T$ duration. The value staked in the PinFi system $\mathcal{S}_{\text{\tiny PinFi}}$ can be expressed concisely as 
\begin{equation}
\mathcal{S}_{\text{\tiny PinFi}} = \dfrac{\mathbb{S}}{T}\cdot [u(t-t_1) - u(t-t_1-T)](t_1+T-t)
\end{equation}
where $u(\cdot)$ is the step function.  This expression illustrates a linear decrease in staked value from the point of liquidity injection ($t_1$) until the end of the commitment period ($t_1+T$), after which the staked value returns to zero, embodying the inherently dissipative nature of liquidity in PinFi systems. Through this model, PinFi systems adeptly manage the transient characteristics of physical computing resources, offering a structured approach to capturing the dynamic interplay between time, liquidity, and computing power.

Through this model, we deduce the dissipation rate in a PinFi system as follows:
\begin{equation}
\mathcal{D}_{\text{\tiny PinFi}} = \dfrac{\mathbb{S}}{T}\cdot[\delta(t-t_1) - \delta(t-t_1-T)](t_1+T - t) -\dfrac{\mathbb{S}}{T}\cdot[u(t-t_1) - u(t-t_1-T)].
\end{equation}
Here $\delta(\cdot)$ is the Kronecker delta function. This dissipation rate is 
\begin{equation}
\mathcal{D}_{\text{\tiny PinFi}} = 
\begin{cases}
0, \ \ \ & t< t_1,\\[6pt]
\mathbb{S}\cdot \delta(t-t_1), \ \ \ & t\rightarrow t_1^-,\\[6pt]
\mathbb{S}\cdot \delta(t-t_1) -\dfrac{\mathbb{S}}{T}, \ \ \ & t\rightarrow t_1^+,\\[6pt]
-\dfrac{\mathbb{S}}{T}, \ \ \ & t\in (t_1, t_1+T^-],\\[6pt]
0, \ \ \ & t\in [ t_1+T^+, + \infty).
\end{cases}
\end{equation}
Key observations regarding the dissipation rate in the PinFi system include:
\begin{itemize}
\item The presence of a non-zero dissipation rate throughout the staking period ($T$), underscoring the intrinsic dissipative nature of PinFi systems compared to their DeFi counterparts.
\item The dissipation rate's non-smooth transitions at the liquidity injection point ($t=t_1$) and at the expiration of the staking period ($t=t_1+T$).
\end{itemize}

This model illustrates how PinFi systems, through their time-limited staking approach, inherently possess a dissipative quality. As the staking duration ($T$) extends indefinitely ($T\rightarrow +\infty$), the behavior of PinFi systems begins to resemble that of traditional DeFi systems. The juxtaposition of dissipative tokens (e.g., computing power hours) against non-dissipative protocol tokens presents unique challenges and considerations for the automatic market maker protocol, highlighting the need for innovative solutions tailored to the dynamics of PinFi ecosystems.

Moreover, the presence of discontinuities in the PinFi system could potentially compromise the stability of the liquidity pool when considering multiple liquidity injection events.  If we assume the $i^{\text{\tiny th}}$ staking event injects a value of $\mathbb{S}_i$ representing computing power hours for a duration of $T_i$ hours starting at time $t_i$, the cumulative staked computing power at any given time tt can be summarized as:
\begin{equation}
\mathcal{S}_{\text{\tiny PinFi}} = \sum_i\dfrac{\mathbb{S}_i}{T_i}[u(t-t_i) - u(t-t_i-T_i)](t_i+T_i-t)
\end{equation}
where $u(\cdot)$ is the step function. 

Given the decentralized and permissionless nature of the protocol, controlling the timing ($t_i$) and duration ($T_i$) of these staking events is infeasible. As discontinuities manifest at $t=t_i$ and $t=t_i+T_i$, they can rapidly propagate, affecting the liquidity pool at any given time $t$. This highlights the necessity for strategic planning and mechanism design within PinFi systems to accommodate and mitigate the effects of these discontinuities, ensuring the long-term viability and stability of the liquidity pool.

\subsection{Implementing PinFi Liquidity Pools: A Practical Approach}

To tackle the challenges arising from the dissipative nature of computing power hours and to maintain a stable exchange ratio within the liquidity pool, two key mechanisms are utilized: a periodic withdrawal mechanism and an periodic renewal mechanism. These mechanisms are designed to dynamically adjust the pool's composition, counteracting the gradual dissipation of staked computing power. Through these adjustments, the liquidity pool's equilibrium and operational functionality are preserved, ensuring a consistent and reliable ecosystem for all participants.

\begin{figure}[ht]
\centering
\includegraphics[width=1.0\linewidth]{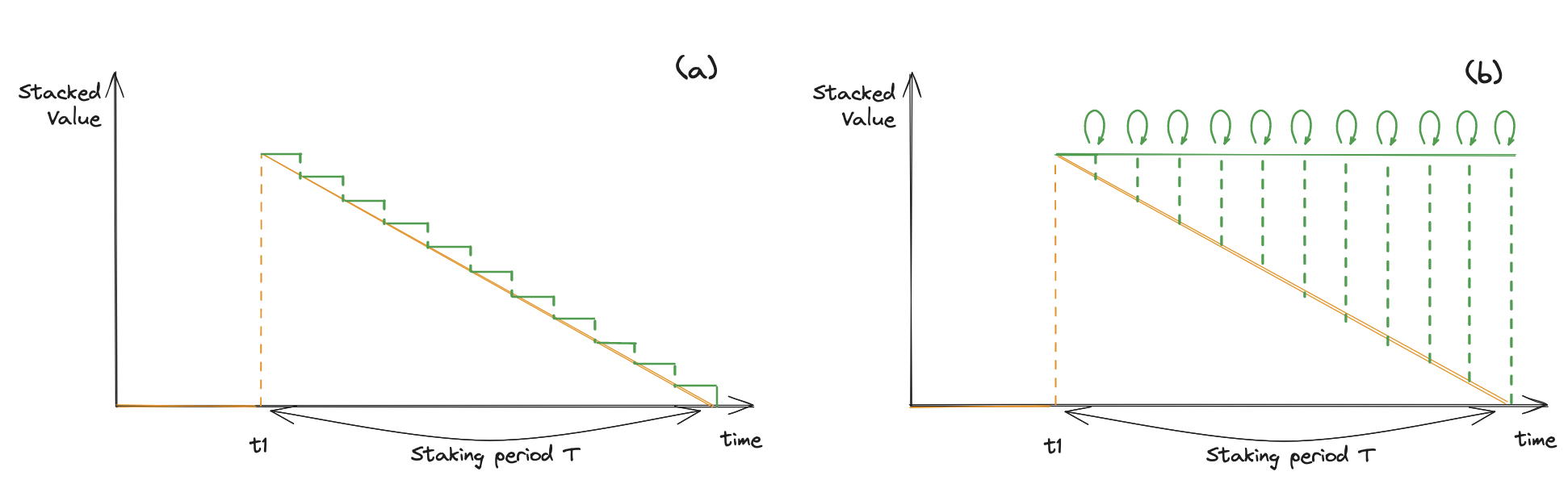}
\caption{\label{fig:praticalf}(a). Periodic withdrawal mechanism. and (b). Periodic renewal mechanism}
\end{figure}

\subsubsection{Periodic Withdrawal Mechanism}

In our practical implementation, we adopt multiple step functions to approximate the original continuous decay behavior of the staked computing power, as illustrated in Figure \ref{fig:praticalf}(a). This approach effectively eliminates the discontinuities observed around the times of liquidity injection and expiration. Additionally, managing the exchange ratio within the liquidity pool becomes more straightforward. By transforming the continuous dissipation of computing-power-hours into a discretely stepped process, we can apply a periodic withdrawal mechanism. This method simplifies the handling of liquidity adjustments over time, ensuring the stability and efficiency of the pool's operations while maintaining the integrity of the exchange ratio.

\subsubsection{Periodic Renewal Mechanism}

In our practical implementation, the smart contract initiates an interaction with the miner at the conclusion of each epoch to inquire if they wish to extend their staking for an additional epoch ($\Delta$). Should the miner consent to the extension and maintain the required maintenance deposit in their address, the protocol token will not be removed and the computing power hour token will be renewed. This adoption of a periodic renewal mechanism effectively transforms our scenario into one that mirrors the traditional DeFi system framework. This approach not only facilitates a more seamless management of liquidity and staking durations but also enhances the flexibility and attractiveness of the PinFi system for miners, aligning it more closely with the established practices of DeFi systems while maintaining the unique characteristics and advantages of the PinFi model.

\section{Transactions and System Balance}
\label{txs}

The LooPIN project offers a comprehensive transactional framework that supports a variety of transfers, including address-to-address, contract-to-address, address-to-contract, and contract-to-contract movements of protocol tokens. This flexibility is akin to the transactional capabilities found in many existing blockchain networks. Beyond these standard transfer types, the LooPIN project introduces a suite of primitives designed specifically to cater to the operational necessities of a PinFi (Physical Infrastructure Finance) protocol.

These primitives are fundamental to enabling the core functionalities that distinguish the PinFi protocol within the decentralized finance (DeFi) landscape, particularly in the realm of computing power networks. They facilitate critical operations such as resource staking, dynamic pricing mechanisms, liquidity management, and secure access to computing resources. 

The LooPIN project incorporates two primary mechanisms for utility token burn, designed to ensure commitment and contribute to the network's stability and security. These mechanisms are as follows:

\begin{itemize}

\item {\bf{Staking Fees}}: Devices participating in the LooPIN network are required to "burn" a certain amount of utility tokens as staking fees. This act serves as a pledge of their resources to the network's liquidity pools. The burning of staking fees is a critical process that signifies the device's commitment to contributing its computing resources to the network. It acts as a gatekeeping mechanism, ensuring that only serious and committed participants can contribute to and benefit from the network's resources.

\item {\bf{Exchange Fees}}: In addition to the initial staking fees, both users seeking to withdraw liquidity and liquidity removal providers must incur a specified exchange fee, which is burned during the process of extracting liquidity. The rationale behind this requirement stems from the liquidity providers' role: their passive contribution of liquidity enables others to sell their ``future services" to users. To acknowledge and incentivize these contributions, new protocol tokens, generated from newly minted blocks, are awarded to liquidity providers as rewards. This mechanism ensures equitable compensation, fostering continuous liquidity provision and supporting the ecosystem's stability and growth.
\end{itemize}
These burn mechanisms play a crucial role in the LooPIN protocol's economic model, regulating the supply of utility tokens while incentivizing the provision and maintenance of high-quality computing resources within the network. By requiring the burning of utility tokens for participation and ongoing engagement, the protocol aligns the interests of device operators with the overall health and security of the network, fostering a robust and reliable ecosystem for decentralized computing power.

A concise analysis offers valuable insights into the parameters' boundaries crucial for maintaining equilibrium within our protocol. For the sake of simplicity, we define the Staking Fee as $\alpha$, the Maintenance Burning Rate as $\beta$, and the Rewarding Rate as $\gamma$. Let $T$ represent the total staking duration for a computing device, and $T_1$ be the point in time when the device experiences a malfunction.
To ensure the system's balance, the following condition must be met:
\begin{equation}
-\alpha -\beta T + \gamma T_1 = 0
\end{equation}
This equation implies that for any given $T_1$ that is less than or equal to $T$, a balance is achievable. Reformulating the equation, we obtain:
\begin{equation}
T_1 = \frac{\alpha}{\gamma} + \frac{\beta}{\gamma}T \leq T
\end{equation}
This leads us to deduce that $\beta\leq \gamma$, ensuring that the Maintenance Burning Rate does not exceed the Rewarding Rate. Additionally, it establishes that $T \geq T_{\text{\tiny min}} = \dfrac{\alpha}{\gamma - \beta}$, where $T_{\text{\tiny min}} = \dfrac{\alpha}{\gamma - \beta}$ represents the minimum required staking period for a device to remain viable within the system. Both imply that the design of the rewarding system must carefully consider these rates to maintain an equilibrium that encourages long-term participation and contribution to the network.

\section{Ongoing Work}
\label{ongoingwork}

This document outlines the foundation of the LooPIN protocol, marking the initial phase in the broader scope of engineering, research, and design efforts aimed at enhancing PinFi within computing power networks. There are several ongoing directions for exploration and development:

\begin{itemize}

\item {\bf{Expanding Scope to Other DePIN Applications}}: We aim to explore the viability of extending the principles underpinning the LooPIN protocol to additional Decentralized Physical Infrastructure Networks (DePIN), such as those governing electrical grids\cite{NourCS22} and file storage systems. This exploration will ascertain the adaptability and effectiveness of our protocol across different infrastructural domains.

\item {\bf{Enhancing Security with Advanced Proofs-of-Computing-Power-Staking}}: As the LooPIN protocol scales, maintaining its security becomes paramount. We plan to research and deploy sophisticated Proofs-of-Computing-Power-Staking mechanisms, ensuring the network's integrity and resilience against potential vulnerabilities.

\item {\bf{Optimizing Liquid Pool Dissipation Rates}}: To provide a seamless user experience and mitigate arbitrage opportunities, we will focus on refining the dissipation process within the liquidity pools. By decreasing the time increments involved, we aim to achieve a smoother and more efficient interaction for all participants.

\item {\bf{Incentive System Game-Theoretic Analysis}}: A thorough game theoretic analysis of the incentive mechanisms within the LooPIN protocol will be conducted. This analysis will help identify equilibrium strategies, ensuring that the incentives align stakeholders' actions towards the protocol's long-term sustainability and success.

\item {\bf{Formal Verification of Scoring and Ranking Algorithms}}: The integrity of the Proof-of-Computing-Power-Staking depends significantly on the scoring and ranking algorithms employed. We commit to rigorously proving these algorithms' correctness and efficiency, thereby solidifying the foundation upon which our protocol stands.
\end{itemize}

These ongoing work represents our commitment to pushing the boundaries of decentralized computing power networks. 
Through research and engineering, we aim to unlock new possibilities for the LooPIN protocol and pave the way for a more decentralized, efficient, and secure digital infrastructure.

\section*{Acknowledgements}
This document stands as a testament to the dedication of our team members, whose tireless collaboration was essential to its realization. 
%The invaluable support, insightful feedback, and thorough review provided by our board of directors, advisors, investors, and collaborators have been pivotal to our success. To each individual involved, we extend our deepest gratitude for their contributions.

\bibliographystyle{alpha}
\bibliography{sample}

\end{document}